\documentclass[%
 reprint,
nofootinbib,
 amsmath,amssymb,
 aps,
]{revtex4-2}

\usepackage[justification=raggedright,singlelinecheck=false]{caption}
\usepackage{subcaption}

\usepackage{graphicx}
\usepackage{dcolumn}
\usepackage{bm}
\usepackage{bbm}
\usepackage{comment}
\usepackage{hyperref}
\usepackage{xcolor}
\usepackage[normalem]{ulem}

\begin{document}

\preprint{APS/123-QED}

\title{Contrastive Learning for Robust Representations of
Neutrino Data}

\author{Alex Wilkinson}
\email{alex.j.wilkinson@warwick.ac.uk}
 \affiliation{%
 University College London, London, WC1E 6BT, United Kingdom
}%
\affiliation{%
 University of Warwick, Coventry, CV4 7AL, United Kingdom.
}%
\author{Radi Radev}%
 \email{radi.radev@meteoswiss.ch}
\affiliation{%
 CERN, The European Organization for Nuclear Research, CH-1211 Meyrin, Switzerland
} 
\affiliation{Federal Office of Meteorology and Climatology MeteoSwiss, CH-8058 Zurich, Switzerland.}
\author{Sa\'ul Alonso-Monsalve}%
 \email{salonso@ethz.ch}
\affiliation{%
 ETH Zürich, Institute for Particle Physics and Astrophysics, CH-8093 Zurich, Switzerland.
}%

\date{\today}

\begin{abstract}
In neutrino physics, analyses often depend on large simulated datasets, making it essential for models to generalise effectively to real-world detector data. Contrastive learning, a well-established technique in deep learning, offers a promising solution to this challenge. By applying controlled data augmentations to simulated data, contrastive learning enables the extraction of robust and transferable features. This improves the ability of models trained on simulations to adapt to real experimental data distributions. In this paper, we investigate the application of contrastive learning methods in the context of neutrino physics. Through a combination of empirical evaluations and theoretical insights, we demonstrate how contrastive learning enhances model performance and adaptability. Additionally, we compare it to other domain adaptation techniques, highlighting the unique advantages of contrastive learning for this field.
\end{abstract}

\maketitle

\section{\label{sec:intro}Introduction}

High-energy physics simulations, which primarily rely on Monte Carlo (MC) methods, constitute the primary source for training machine learning (ML) algorithms in particle physics. The reason is that they provide ``labelled'' data and allow for the generation of large sample sizes at an affordable computational cost. However, simulations are generally far from ideal, with imperfections arising from uncertainties in theoretical models, approximations in the physics processes, and detector effects that are not fully accounted for in the simulations. As a result, detector data and MC simulations often come from different underlying distributions~\cite{DANIELELVIRA20171, nachman2019aisafetyhighenergy}.

Neutrino physics is no exception to this practice, as it also relies heavily on simulations, where some unknowns in the model must be parametrised~\cite{coyle2022impact, PhysRevD.106.032009, ershova2023role}. To address these challenges, it is essential to design models that are invariant to these parameters where possible and to characterise the dependence of the models on these parameters when invariance cannot be achieved. Moreover, there will inevitably be ``unknown unknowns'' (i.e., systematic effects or mismodellings that are not anticipated), and efforts can still be made to identify ways in which detector data can be leveraged to mitigate the impact of these unknowns on model performance.

In contrast to collider experiments, simulated samples in neutrino oscillation experiments typically consist of data in the form of images, representing the signature of the particles produced by a neutrino interaction in the detector; thus, the ML algorithms preferred for analysing such data primarily come from the family of computer vision techniques~\cite{Aurisano-et-al-2016-convolutional, dunecvn, alonso2024deep}. In this paper, we study the application of contrastive learning of visual representations, based on the SimCLR framework~\cite{chen2020simpleframeworkcontrastivelearning}, to tackle the data-MC discrepancy in neutrino physics. Physics inspired transformations and variations of detector systematic uncertainties are applied to realistic simulations of neutrino detectors to model the presence of distinct data-MC distributions. We employ a contrastive pretraining stage to learn robust representations of the data followed by a task-specific fine-tuning with labelled simulation.

The rest of the paper is organised as follows: Section~\ref{sec:contrastive} provides an overview of contrastive learning and the state-of-the-art, while Section~\ref{sec:method} describes the methodology followed. Section~\ref{sec:datasets} describes the datasets used, and Section~\ref{sec:augmentations} details the augmentations applied. Section~\ref{sec:results} presents the experimental results, and Section~\ref{sec:conclusion} summarises the main findings and outlines directions for future work.

\section{\label{sec:contrastive}Contrastive learning}

One of the primary difficulties in ML lies in training models capable of robust generalisation to unseen data, particularly when originating from distributions distinguishable from the training data~\cite{10.1162/153244302760200704, xu2012robustness, bay2024machine}. This challenge pervades various domains, including neutrino physics, where models are typically trained via supervised learning on extensive simulated datasets and subsequently assessed on unlabelled experimental events from physical detectors~\cite{doi:10.1142/S0217751X20430058, karagiorgi2022machine}. Traditional methods for achieving generalisation entail employing regularisation techniques (e.g., weight decay, dropout, or normalisation layers) and figuring out an optimal balance between training data volume and model complexity~\cite{goodfellow2016regularization, kukacka2017regularization}. However, while explicit regularisation proves beneficial, it may not suffice, leaving models prone to poor generalisation when confronted with target distributions divergent from the training one~\cite{10.1145/3446776, 6135800}. 

Domain adaptation techniques aim to improve the generalisation of models by transferring knowledge from a well-labelled source domain to a target domain characterised by limited or absent labelled data~\cite{NIPS2006_b1b0432c, 10.1007/978-3-030-71704-9_65, csurka2017domain}. These methods can be broadly categorised into feature-based approaches, which align the feature distributions between the two domains, and adversarial methods, which use domain classifiers to learn domain-invariant representations. Techniques such as domain-adversarial training~\cite{dann}, maximum mean discrepancy~\cite{10.1093/bioinformatics/btl242}, and other regularisation-based methods have shown success in reducing domain gaps. A more abstract and complementary paradigm is meta-learning, enabling models to learn how to adapt quickly to new tasks or domains with minimal data. By learning meta-knowledge across tasks, meta-learning techniques can further enhance domain adaptation strategies, particularly in scenarios with limited labelled data from the target domain~\cite{9428530, TIAN2022203, singh2023meta}.

More recently, contrastive learning has regained significant attention as a powerful approach for domain adaptation by leveraging data augmentations and similarities between data points to learn robust, transferable representations \cite{pmlr-v162-shen22d, Thota_2021_CVPR, 10.21468/SciPostPhys.12.6.188}. By encouraging the alignment of similar samples while separating dissimilar ones, contrastive learning is particularly well-suited for addressing discrepancies between domains, such as the data-MC mismatch in high-energy and neutrino physics. Recent methods like RS3L~\cite{PhysRevD.111.032010} use re-simulation to generate physics-driven augmentations, improving model robustness. Large-scale pretraining with contrastive learning has also been explored for collider data, demonstrating gains in adaptability and efficiency~\cite{zhao2024largescalepretrainingfinetuningefficient}.

In the early days of contrastive learning, a siamese architecture was employed to generate embedding vectors $E_{x_{i}}$ and $E_{x_{j}}$ from two inputs $x_{i}$ and $x_{j}$, respectively. A contrastive loss~\cite{1467314} was then used to minimise the distance between $E_{x_{i}}$ and $E_{x_{j}}$ if they belonged to the same class, and to maximise it otherwise. This methodology was later extended with the triplet loss~\cite{7298682}, where, instead of working with pairs of embeddings $(E_{x_{i}}, E_{x_{j}})$ that could belong to the same class or not, each example $x$ was accompanied by a positive sample $x^{+}$ (from the same class as $x$) and a negative sample $x^{-}$ (from a different class than $x$). The triplet loss simultaneously minimises the distance between the embeddings of $x$ and $x^{+}$ while maximising the distance between the embeddings of $x$ and $x^{-}$.

As the field of contrastive learning progressed, these earlier methods were refined to improve scalability and training efficiency. Traditional pair-based and triplet-based methods required careful sampling of positive and negative pairs, which could become computationally expensive and suboptimal when applied to large datasets. To address these challenges, more recent approaches introduced large-scale contrastive frameworks that rely on data augmentations and large batches to construct positive and negative pairs dynamically. The approach used in this paper is SimCLR~\cite{chen2020simpleframeworkcontrastivelearning}, a self-supervised contrastive learning method that removes the need for explicit labels by leveraging data augmentations to generate positive pairs.

In SimCLR, positive pairs are created by applying different augmentations (e.g., random rotations, cropping, or colour distortion) to the same input $x$, while all other samples in the batch act as negative pairs. A contrastive loss, such as the NT-Xent loss~\cite{NIPS2016_6b180037}, is used to bring the embeddings of positive pairs closer together while pushing the embeddings of negative pairs apart. This approach significantly simplifies contrastive learning by leveraging the diversity of augmentations and large-batch training, enabling the learning of high-quality representations without the need for explicit class labels.

These developments have contributed to the rise of self-supervised learning, which trains models on large, unlabelled datasets to extract generalisable features. Frameworks such as SimCLR, MoCo~\cite{he2020momentumcontrastunsupervisedvisual}, or BYOL~\cite{10.5555/3495724.3497510} exemplify methodologies that allow models to learn visual representations in a contrastive manner. When trained on sufficiently diverse and large datasets, models obtained through such methodologies can serve as foundation models capable of being fine-tuned for a variety of downstream tasks, including domain adaptation in experimental physics.
~


\section{\label{sec:method}Method}

 \begin{figure*}[tbh]
   \includegraphics[width=1.0\textwidth]{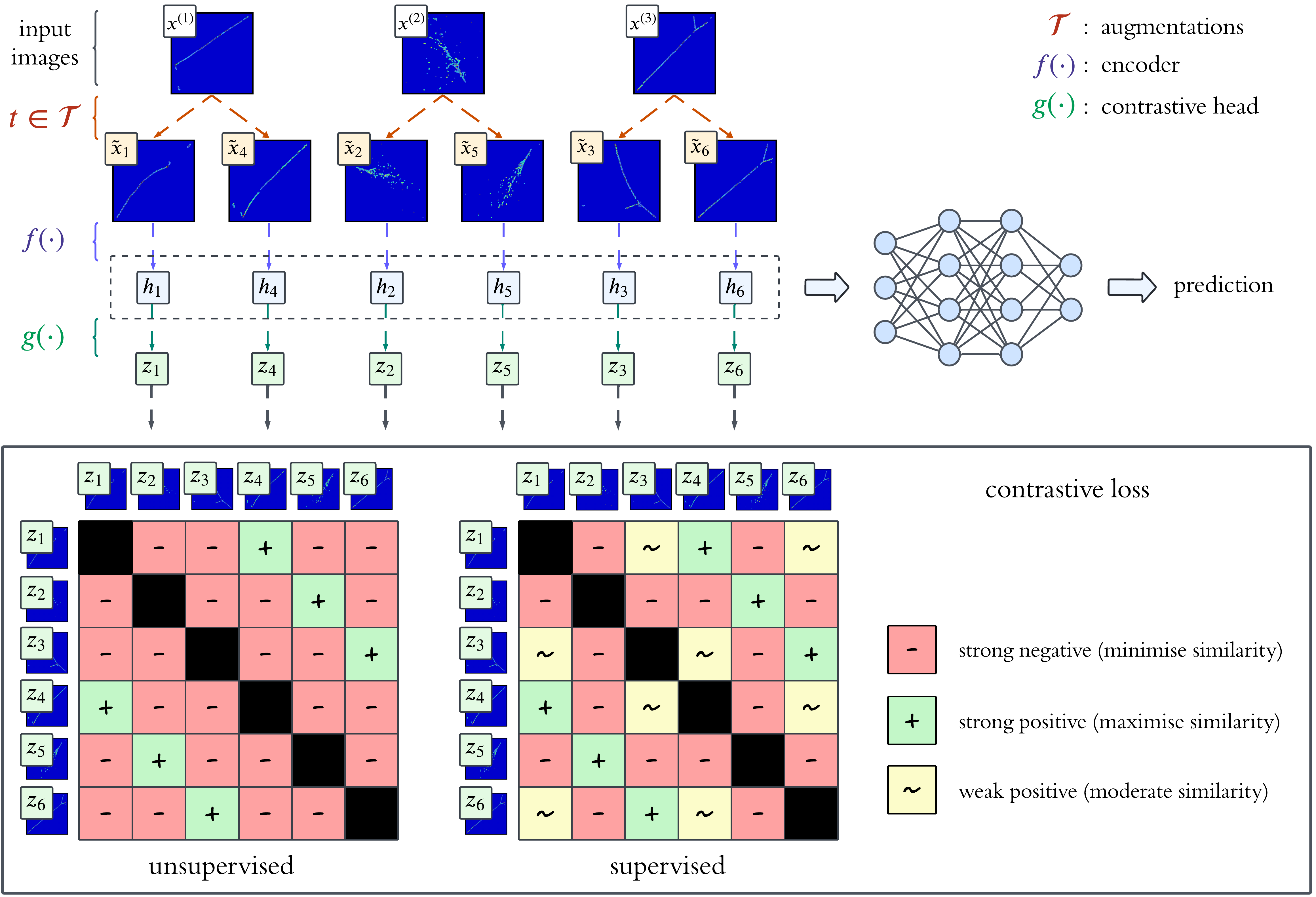}
   \caption{Diagram illustrating the method for contrastive learning applied to neutrino images. The approach involves augmentations $\mathcal{T}$ of input images, processing through an encoder $f(\cdot)$ that maps the inputs to a latent representation space $h_i$, and a contrastive head $g(\cdot)$ that further transforms these representations $h_i$ into $z_i$ for the contrastive objective. In unsupervised contrastive loss, pairs are formed solely based on augmented views of the same data point, without relying on class labels. In contrast, supervised contrastive loss leverages labels to guide the formation of positive (same event), moderate (different event, same class), and negative pairs (different class), enabling improved representation learning. The latent representations $h_i$ are then used as robust inputs for a classifier neural network.}
   \label{fig:method}
\end{figure*}
We consider unsupervised and supervised contrastive learning approaches for the task of reconstruction on realistic high-resolution 3D neutrino detector simulation. These approaches enable an encoder network $f_\theta(.)$ to learn a robust representation $h$ of the data, captured in its weights $\theta$. These representations $h$ can then be leveraged for downstream tasks by fine-tuning the network using labelled data (Figure~\ref{fig:method}).

Although our method generalises to various input modalities, in this work, each event (example) $x$ is represented by two tensors: a coordinate tensor $c \in \mathbb{R}^{N \times 3}$ containing the 3D spatial locations of $N$ voxels, and a feature tensor $v \in \mathbb{R}^{N \times M}$ containing their corresponding values across $M$ modalities. Our trained network ensures that our extracted representation remains robust, i.e., $f(c, v) \approx f(\mathcal{T}(c,v))$, where $\mathcal{T}$ represents the set of augmentations applied.

To fine-tune the network, we first freeze the backbone $f_\theta(.)$ pretrained during the contrastive learning; then we train a small classifier on the features extracted from this backbone - we found the choice of architecture for this classifier to have minimal impact on performance. In all experiments, we used logistic regression implemented using \texttt{LogisticRegression} from \texttt{scikit-learn} \cite{scikit-learn}.

\subsection{Architecture}
Neutrino data - especially collected from a 3D detector, such as pixelated liquid argon time projection chamber (LArTPC) \cite{lartpc_detector} and segmented scintillators \cite{blondel2020superfgd} considered in this work, is extremely sparse \cite{adams2020pilarnet}. Naively applying conventional deep learning methods that rely on 3D convolutions or other gridded methods is infeasible due to the immense memory and compute requirements to process the high-resolution data (e.g., $500^3$ voxels) present in these detectors. 

When processing highly sparse, high-resolution neutrino detector data, three primary approaches emerge: graph-based methods, point-cloud representations, and sparse voxel grids. Neutrino graph-based approaches present ambiguity when defining the connectivity in the initial graph. Point-cloud approaches require transforming the naturally voxelised detector data into point representations, thus introducing unnecessary complexity. The inherent voxelised structure of data acquisition in segmented scintillator and LArTPC detector systems naturally favours a voxelised representation over alternatives. This alignment with detector readout, combined with demonstrated success across multiple experiments, has established sparse voxel methods as the standard approach in neutrino physics deep learning applications \cite{domine2020scalable, drielsma2021scalable, PhysRevD.105.112009, microboone_sparse}.  

We base our encoder architecture on ConvNeXt V2 \cite{woo2023convnextv2}. ConvNeXt V2 was originally designed as a masked asymmetric autoencoder, where the decoder is discarded after pre-training and a classification head is fine-tuned; however, in our approach, we sacrifice its autoencoder nature and instead employ contrastive learning during pre-training. During the training of the contrastive learning model an MLP is added to compute the feature similarity for the contrastive losses - Equations \ref{eqn: simclr} and \ref{eqn: sup_contrastive}. Following the SimCLR methodology, this MLP is removed during the fine-tuning phase. 

All our models are implemented using a custom version~\footnote{The version of MinkowskiEngine used includes a custom CUDA kernel for depth-wise convolutions~\cite{Chollet_2017_CVPR}: \url{https://github.com/shwoo93/MinkowskiEngine}. The docker image used can be found here: \url{https://hub.docker.com/r/rradev/minkowski:torch1.12_final}.} of \texttt{MinkowskiEngine} - a deep learning framework built on top of \texttt{Pytorch} with efficient kernels for processing highly sparse voxelised data \cite{choy20194d, paszke2019pytorch}. The code used to generate the results presented in Sec.~\ref{sec:results} is available at: \url{https://github.com/radiradev/contrastive-neutrino}. 

Using our encoder backbone, we consider unsupervised and supervised versions of contrastive learning \cite{chen2020simpleframeworkcontrastivelearning, khosla2020supervisedcontrastive}. The same encoder backbone with a learned representation $h$ of dimension 768 is used for all architectures considered in our experiments. Training is performed with a large batch size of 672 to maximise the number of negative pairs in each iteration.

\subsection{\label{sec:unsupervised} Unsupervised}

During the pre-training phase, unsupervised contrastive learning as in SimCLR \cite{chen2020simpleframeworkcontrastivelearning} is trained purely on data without labels, and the labels are only used during the fine-tuning phase. Formally, the loss function is defined as:
\begin{equation}
\mathcal{L}_{\text{SimCLR}} = -\log \frac{\exp(\text{sim}(\mathbf{z}_i, \mathbf{z}_j)/\tau)}{\sum_{k} \mathbbm{1}_{[k \neq i]} \exp(\text{sim}(\mathbf{z}_i, \mathbf{z}_k)/\tau)},
\label{eqn: simclr}
\end{equation}
where \(\text{sim}(\mathbf{z}_i, \mathbf{z}_j)\) denotes the cosine similarity between representations \(\mathbf{z}_i\) and \(\mathbf{z}_j\), and \(\tau\) is a temperature parameter. This loss is aggregated over all positive pairs \((i,j)\) in a batch.

Since labels are not utilised in the pre-training phase, it can be applied to data from the target distribution. This ensures the features learned  by the encoder network are not particular to the training distribution used for fine-tuning, facilitating a domain adaptation.

\subsection{\label{sec:supervised} Supervised}

To incorporate class information into the contrastive loss, a supervised contrastive loss function can be employed \cite{khosla2020supervisedcontrastive}. This extends the SimCLR loss by utilising label information to group samples of the same class while separating samples of different classes. The supervised contrastive loss for a given anchor sample \(i\) is defined as:
\begin{equation}
\mathcal{L}_{\text{sup}} = -\log \frac{\sum_{j} \mathbbm{1}_{[y_i = y_j]} \alpha \exp\left(\text{sim}(\mathbf{z}_i, \mathbf{z}_j)/\tau\right)}{\sum_{j} \exp\left(\text{sim}(\mathbf{z}_i, \mathbf{z}_j)/\tau\right)},
\label{eqn: sup_contrastive}
\end{equation}
where \(y_i\) and \(y_j\) denote the labels of samples \(i\) and \(j\), and \(\alpha\) is a weighting factor balancing contributions from same-class positives with same-image positives (set to 1 in our experiments). The numerator aggregates contributions from all same-class positive pairs of the anchor sample, weighted by \(\alpha\), while the denominator considers all pairs of the anchor sample. This loss is aggregated over all anchor samples \(i\) in the batch.

By leveraging label information, the supervised contrastive loss explicitly encourages intra-class similarity and inter-class separation, providing a more structured representation for downstream tasks.

\subsection{\label{sec:dann} Extra: domain adversarial neural networks}

To compare our results with other domain adaptation techniques, we selected the domain adversarial neural network (DANN)~\cite{dann} due to its established applications in neutrino physics~\cite{Perdue_2018, PhysRevD.105.112009}. DANN is a domain adaptation method that aims to minimise the discrepancy between the source and target domains by employing adversarial training. The architecture of DANN includes three main components: a feature extractor, a label predictor, and a domain classifier. The feature extractor learns a representation of the input data, while the label predictor is trained on labelled source data to predict task-specific labels - here, it shares the same architecture with the same backbone encoder used for contrastive cases. Simultaneously, the domain classifier is trained to distinguish between source and target domains. By applying a gradient reversal layer between the feature extractor and the domain classifier, the feature extractor is encouraged to generate domain-invariant features. This adversarial process aligns the feature distributions of the source and target domains, enabling effective domain adaptation.

\section{\label{sec:datasets}Simulated datasets and augmentations}

\subsection{\label{sec:lartpc}Pixelated LArTPC Detector}

The dataset simulates the readout of a pixelated LArTPC detector. There are five simulated particle types (protons, pions, muons, photons, and electrons). These particles are simulated uniformly within the detector volume. All particle types except protons have an energy uniformly selected from the range [0.05, 1.00] GeV. Protons have an energy range of [0.05, 0.40] GeV. Realistic particle propagation within the detector is simulated using \texttt{edep-sim} \cite{edep}, a \textsc{Geant4} \cite{GEANT4:2002zbu} wrapper, and the electronics response of the pixelated detector is simulated using \texttt{larnd-sim} \cite{larndsim}. Each event in the dataset $x$ is a tensor $v \in \mathbb{R}^{N}$ containing ADC counts (related to the deposited energy of the detected particle) and their corresponding coordinates $c \in \mathbb{R}^{N \times 3}$. 

The code to generate this dataset is available at: \url{https://github.com/radiradev/larpix-detector-sim}.

\subsection{\label{sec:plastic_scintillator} Segmented Scintillator Cube Detector}

The dataset consists of simulated data for a three-dimensional fine-grain plastic scintillator detector with a size of \(2 \times 2 \times 2~\mathrm{m^3}\). The detector is divided into \(200 \times 200 \times 200\) cubic voxels of \(1~\mathrm{cm^3}\) each, where each voxel acts as a sensor for scintillation light produced by traversing particles. Single-charged particles (protons, pions, muons, and electrons) with isotropic directions and uniform momentum distributions are simulated within the detector volume. Detector response simulation accounts for crosstalk, the leakage of scintillation light from one voxel into adjacent voxels, modelled with a 3\% leakage probability per face. Each event in the dataset $x$ is a tensor $v \in \mathbb{R}^{N}$ containing the number of detected photoelectrons per voxel (cube) and their corresponding coordinates $c \in \mathbb{R}^{N \times 3}$. This publicly available dataset can be accessed at: \url{https://doi.org/10.5281/zenodo.10998285}.

\section{\label{sec:augmentations}Augmentations}

We use a series of augmentations during the training of the contrastive learning models, as well as when training a baseline classifier with augmentations, which refers to a standard supervised classifier trained with data augmentations applied to the input samples during training. This ensures that the classifier is exposed to a range of transformations similar to those used in the contrastive learning setup, allowing a direct comparison of their effects on robustness to domain shifts. We apply the same augmentations to both datasets. These datasets were augmented using a series of transformations designed to introduce spatial and energy variations, simulating realistic conditions and improving model robustness. During training, for both the contrastive models and the classifier with augmentations, we randomly sample with replacement a series of three augmentations from a full set of augmentation transformations. The augmentations applied include: a rotation transformation that randomly rotates the spatial coordinates; a masking operation that randomly drops a fraction of the input points, effectively simulating partial data loss; an energy scaling augmentation that introduces voxel-wise energy shifts by scaling the feature values with random noise sampled from a normal distribution; a spatial translation transformation that displaces the input points by a random offset proportional to a predefined scale factor; and the identity transformation. These augmentations ensured compatibility with batched processing and scalable data handling.

We also experimented with augmentation via re-simulation of the detector response for the LArTPC dataset, similar to approaches explored in high-energy physics~\cite{PhysRevD.111.032010}. We re-simulated events at the \texttt{larnd-sim} level, altering parameters related to the electric field, transverse and longitudinal diffusion and electron lifetime. However, when using these augmentations with realistic simulation parameter ranges, we found the augmentations to be too weak. In line with previous work (making the case for using \textit{strong} augmentations)~\cite{chen2020simpleframeworkcontrastivelearning}, our trained contrastive model using purely re-simulation was unable to produce any useful features for downstream tasks. We also tried using re-simulation augmentations to create additional augmentations to complement the ones described in the previous paragraph. This resulted in marginal gains even when evaluated on samples that had undergone re-simulation with the same altered parameters as used in the augmentations. Given the significant computational requirement, we advocate against the use of re-simulation for augmentations in this use case.

\begin{figure}[h!] 
    \centering
    \begin{subfigure}[t]{0.15\textwidth}
        \includegraphics[width=\textwidth, trim=50 50 40 50, clip]{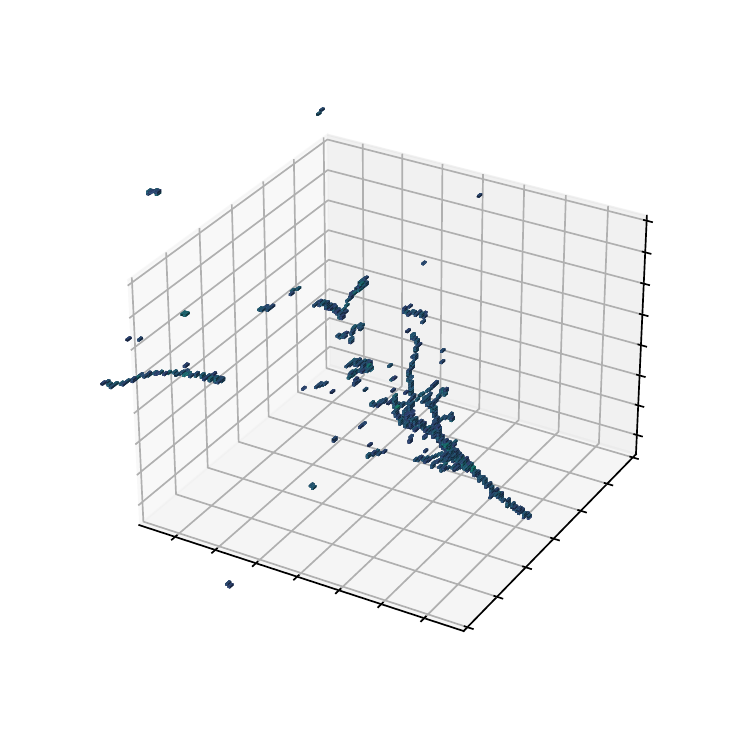}
        \caption{Nominal}
    \end{subfigure}%
    ~
    \begin{subfigure}[t]{0.15\textwidth}
        \includegraphics[width=\textwidth, trim=50 50 40 50, clip]{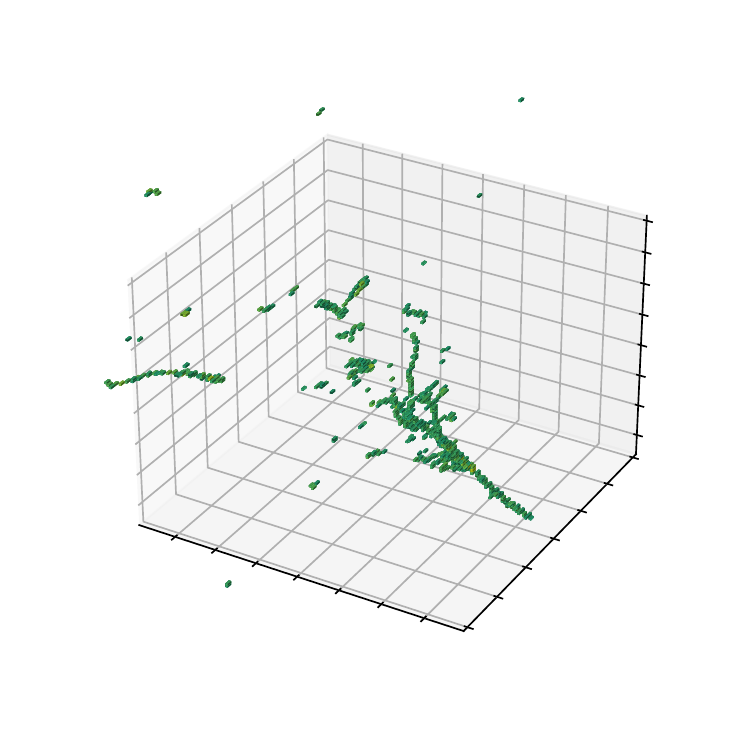}
        \caption{Throw 1}
    \end{subfigure}%
    ~
    \begin{subfigure}[t]{0.15\textwidth}
        \includegraphics[width=\textwidth, trim=50 50 40 50, clip]{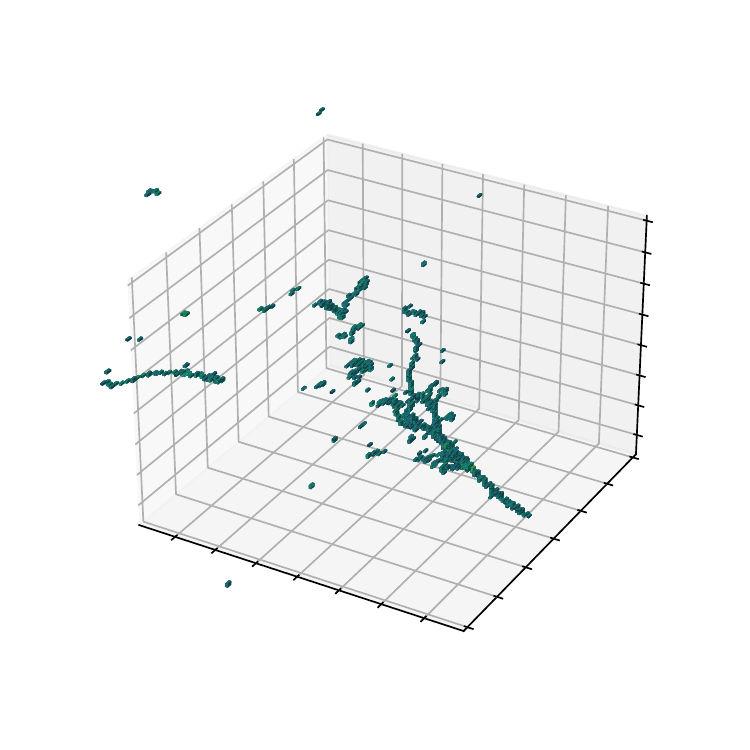}
        \caption{Throw 2}
    \end{subfigure}
    \caption{Examples of thrown electronics response parameters for an electron event of the LArTPC dataset. The effect is a composition of ADC scaling, dropping of voxels, and shifting of voxel positions.}
    \label{fig:elec_throws}
\end{figure}

\section{\label{sec:results}Experimental results}

 \begin{figure*}[tbh]
   \includegraphics[]{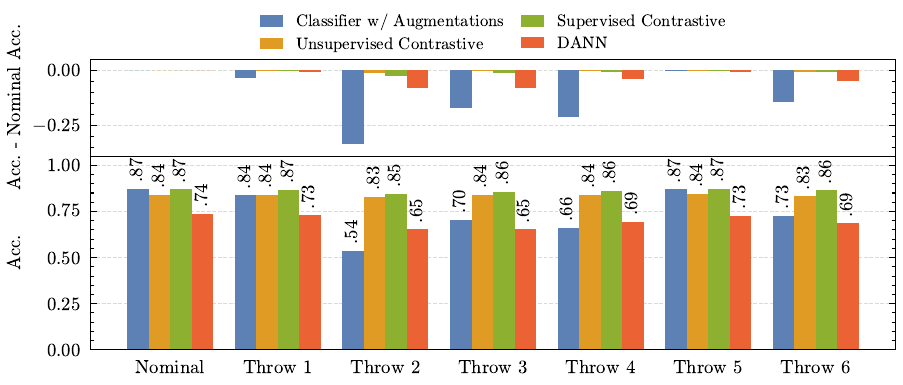}
   \caption{Bar chart of test accuracy of our methods - unsupervised and supervised contrastive learning, compared to a DANN and a classifier with training data augmentations. Nominal indicates the test accuracy of the models using the same detector simulation parameters used in training (i.e., no distribution shift). During testing, these methods are evaluated on various data throws - shifts in detector simulation parameters chosen randomly as shown in Table \ref{table:parameter_throw_variations}.}
   \label{fig:bar_chart}
\end{figure*}

 \begin{figure*}[tbh]
   \includegraphics[]{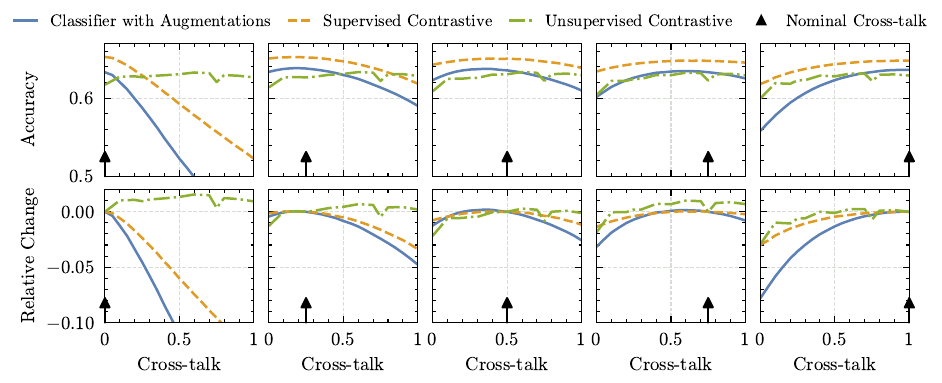}
   \caption{Top: Accuracy of our proposed methods - supervised and unsupervised contrastive learning compared to a classifier trained with augmentations on the scintillator detector dataset. Bottom: Relative change from nominal (training) level of cross-talk to the current (testing) cross-talk level. Nominal cross-talk is the level used to train the classifier and fine-tune the contrastive learning methods. The models are evaluated on cross-talk increments of 0.05-0.10, ranging from no cross-talk to the maximum level of 1.00.}
   \label{fig:seg_cube}
\end{figure*}

For the LArTPC dataset, the trained models were evaluated on several \textit{throws}, datasets simulated using detector parameters that have been shifted from the nominal values used to generate the training dataset. The throws are intended to emulate data from a target distribution that is distinct from the training distribution. In Table~\ref{table:parameter_throw_variations}, we present the detector simulation parameters that were varied to produce the thrown evaluation datasets. The parameters represent uncertainty in the electronic response of the detector and in several noise sources. The parameters for a throw are randomly sampled from a normal distribution centred on the nominal value and with the width listed in the table. An example of the effect of such throws is shown in Figure~\ref{fig:elec_throws}. The altered detector simulation parameters have a significant effect on superficial properties of the event while leaving the topology largely unchanged.  Figure~\ref{fig:bar_chart} shows, for the LArTPC dataset, how the accuracies of the considered methods are affected by varying the detector parameters shown in Table \ref{table:parameter_throw_variations}.

\begin{table}
\caption{Parameter variations used in the simulation of detector response. Throw 1$\sigma$ shows the ($\pm$) percentage variation from the nominal value at one standard deviation.}
\begin{tabular}{lr}
\hline
Parameter & Throw 1$\sigma$ \\
\hline
Gain & 2\% \\
Buffer Risetime & 10\% \\
Common-mode Voltage & 2\% \\
Reference Voltage & 2\% \\
Pedestal Voltage & 20\% \\
Reset Noise & 10\% \\
Uncorrelated Noise & 10\% \\
Discriminator Noise & 10\% \\
Discriminator Threshold & 2\% \\
\hline
\end{tabular}
\label{table:parameter_throw_variations}
\end{table}

We evaluated the robustness of four different approaches: a classifier with augmentations, unsupervised contrastive learning, supervised contrastive learning, and DANN. Under nominal conditions, the classifier with augmentations achieved the highest performance at 87.3\%, followed closely by our proposed method based on supervised contrastive learning (87.0\%). The unsupervised contrastive learning approach resulted in a 3./\% drop in accuracy due to its inability to utilise the label information present in the dataset during the pre-training phase. The DANN approach led to a substantial performance decline, reaching only 74\%, primarily due to the challenges in the adversarial training and the limited availability of diverse training examples for each throw.

The supervised contrastive method exhibited remarkable stability across all parameter variations, maintaining accuracy within the 84.5--87.3\% range. In contrast, the classifier with augmentations demonstrated significant sensitivity to parameter variations, particularly in Throw2, where its performance dropped sharply to 54\%, corresponding to a 33\% decrease from its nominal performance. The unsupervised contrastive approach showed consistent performance across all variations, with accuracies ranging from 83.0--84.1\%. DANN consistently performed below the other methods, with accuracies ranging from 65.3--73.5\%, indicating robustness but at the cost of significantly reduced accuracy across all parameter variations.

In Figure \ref{fig:seg_cube}, we summarise the results of our proposed method on the scintillator detector dataset. Based on supervised and unsupervised contrastive learning, our methods are compared against a classifier using the same architecture. In Appendix \ref{sec:appendix}, we also compare the classifier without augmentations and a DANN. 

The classifier is trained with the same augmentations applied in the contrastive learning pre-training setup. We generate datasets with different levels of cross-talk by masking out fractions of the detector response induced by cross-talk in the simulation. We train separate models for each cross-talk level and evaluate these models across all cross-talk levels. The supervised contrastive learning model consistently outperforms all other models except when the models are trained without any cross-talk. The unsupervised contrastive model exhibits almost no degradation as the data domain shifts (changes in cross-talk) during test time. Both contrastive setups maintain superior accuracy (compared to the classifier) as the cross-talk level varies, highlighting that the extracted features depend more on the fundamental physical properties of the particle interactions generating the events. In contrast, the classifier-based approach appears less robust, potentially indicating that its features are more sensitive to specific properties of the detector simulation.

These results strongly suggest that contrastive learning approaches provide enhanced stability against detector simulation variations compared to traditional deep learning classification methods. Looking at Equation \ref{eqn: sup_contrastive}, in this study, we have only considered setting $\alpha$, the factor controlling the weight of same-image versus same-class positives, to 0 or 1, corresponding to fully unsupervised and supervised contrastive learning. Depending on the task, other values may be considered, allowing us to prioritise robustness versus accuracy in the training data domain. The study of both unsupervised and supervised approaches indicates that the contrastive learning approach, rather than only its facilitation of pre-training on an unlabelled target distribution as in the unsupervised case, results in representations robust to distribution shifts. This increased stability is particularly valuable in high-energy physics and related domains, where minor variations in detector conditions can significantly impact model performance. By leveraging contrastive learning, we can ensure that learned representations generalise more effectively across varying conditions, thereby reducing the risk of model performance degradation due to distribution shifts.

\section{\label{sec:conclusion}Conclusion}

In this study, we applied contrastive learning methods to tackle the challenge of differences between simulated and real detector data in neutrino physics. Our results show that contrastive learning improves model robustness against changes in detector conditions, maintaining stable and high performance even when parameters vary. This approach outperformed traditional methods, such as classifiers with augmentations and DANNs. We found that robust representations can be generated with contrastive learning using scalable and generic augmentations and without requiring pre-training on unlabelled data from the target distribution. 

The success of contrastive learning in creating flexible and reliable features makes it a valuable tool for experimental physics, especially in fields where uncertainties in modelling are common. Future research could explore how to combine contrastive learning with other advanced techniques and expand its use to larger datasets or other areas of particle physics. 

Overall, our findings highlight contrastive learning as a promising strategy for building models that perform well under complex and uncertain conditions.

\begin{acknowledgments}
We would like to thank Cristóvão Vilela and Stephen Dolan for their feedback and ideas during early stages of this work. 

This research used resources of the National Energy Research Scientific Computing Center (NERSC), a Department of Energy Office of Science User Facility using award GenAI@NERSC-ERCAP0030533.
\end{acknowledgments}

\appendix
\section{\label{sec:appendix}Results for all Models}

\begin{figure*}[tbh]
   \includegraphics[]{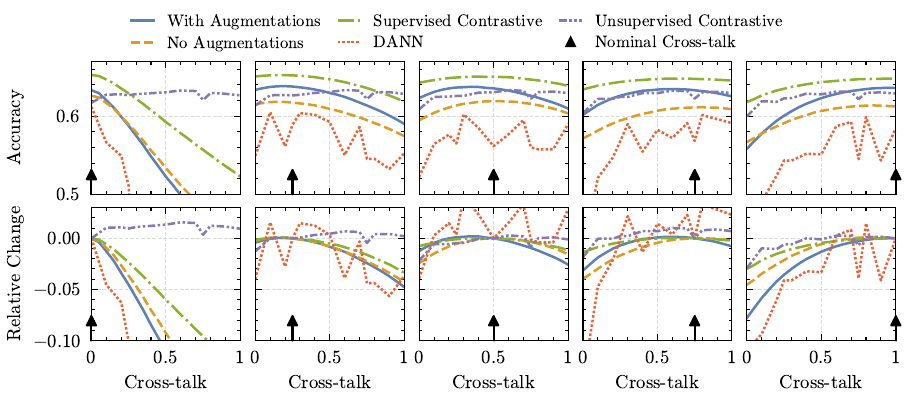}
   \caption{Top: Accuracy of our proposed methods - supervised and unsupervised contrastive learning on the scintillator detector dataset. Our methods are compared to a DANN, and classifiers are trained with and without augmentations. Bottom: Relative change from nominal (training) level of cross-talk to the current (testing) cross-talk level. Nominal cross-talk is the level used to train the classifiers, fine-tune the contrastive learning methods, and provide the labelled source data for the DANN. The models are evaluated on cross-talk increments of 0.05-0.10, ranging from no cross-talk to the maximum level of 1.00.}
   \label{fig:seg_cube_all_models}
\end{figure*}

In Figure \ref{fig:seg_cube_all_models}, we present the results for all trained models on the scintillator detector dataset. Here, we include all models discussed in the main text in Figure \ref{fig:seg_cube} with the addition of a classifier trained without any augmentations and a Domain-Adversarial Neural Network (DANN). The DANN model demonstrates limited accuracy and exhibits substantial performance variability. In contrast, the contrastive learning-based models consistently achieve superior performance and demonstrate enhanced robustness to variations encountered during test time.

\bibliography{apssamp}

\end{document}